\documentclass[11pt]{article}
\usepackage{float}

\usepackage{geometry}                % See geometry.eps to learn the layout options. There are lots.
\usepackage{fullpage}
\usepackage{amsmath,amsfonts,mathrsfs,array,pifont,mathrsfs}
\usepackage{amsthm}
\usepackage{mathrsfs}
\usepackage{graphicx}
\usepackage{graphics}
\usepackage{amssymb}
\usepackage{epstopdf}
\usepackage{epsfig}
\usepackage{times}
\usepackage{subfigure}
\usepackage{hyperref}
\usepackage{color}
\usepackage{yfonts}
\usepackage{cancel}
\usepackage{latexsym}
\usepackage{cite}
\usepackage{multirow}
\usepackage{url}
\usepackage{algorithmic}
\usepackage{mathtools}
\usepackage{stmaryrd}
\usepackage{bm}
\newcommand{\be}{\begin{equation}}
\newcommand{\ee}{\end{equation}}
\newcommand{\bea}{\begin{eqnarray}}
\newcommand{\eea}{\end{eqnarray}}

\newcommand{\de}{{\rm{d}}}

\def\bm1{\mbox{\boldmath $1$}}

\def\e{{\rm e}}

\def\ljump{\mbox{$[ \! [$}}
\def\rjump{\mbox{$] \! ]$}}

\def\calF{\mbox{${\cal F}$}}

\def\calL{\mbox{${\cal L}$}}

\def\calS{\mbox{${\cal S}$}}

\def\be{\begin{equation}}
\def\ee{\end{equation}}
\def\ba{\begin{array}}
\def\ea{\end{array}}
\begin{document}

\title{
A One-Dimensional Peridynamic Model of Defect Propagation\\ and its Relation to Certain Other Continuum Models\\
}

\author{Linjuan Wang$^{1,2}$ and Rohan Abeyaratne$^2$\footnote{Author for
correspondence.}\\[2ex]
$^1$State Key Laboratory for Turbulence and Complex Systems\\
Department of Mechanics and Engineering Science\\
College of Engineering\\
Peking University\\
Beijing 100871, P.R. China\\
wlj@pku.edu.cn\\[2ex]
$^2$Department of Mechanical Engineering\\
Massachusetts Institute of Technology\\
Cambridge, MA 02139, USA\\
rohan@mit.edu\\[2ex]
}

\renewcommand{\thefootnote}{\arabic{footnote}}
\setcounter{footnote}{0}

\maketitle

\vspace{-0.4 in}

\begin{abstract}

The peridynamic model of a solid does not involve spatial gradients of the displacement field and is therefore well suited for studying defect propagation.  Here, bond-based peridynamic theory is used to study the equilibrium and steady propagation of a lattice defect -- a kink -- in one dimension. The material
transforms locally, from one state to another, as the kink passes through. The kink is in equilibrium if the applied force is less than a certain critical value that is calculated, and propagates if it exceeds that value. The kinetic relation giving the propagation speed as a function of the applied force is also derived.

In addition, it is shown that the dynamical solutions of certain differential-equation-based models of a continuum are the same as those of the peridynamic model provided the micromodulus function is chosen suitably.  A  formula for calculating the micromodulus function of the equivalent peridynamic model is derived and illustrated.
This ability to replace a differential-equation-based model with a peridynamic one may prove useful when numerically studying more complicated problems such as those involving multiple and interacting defects.

\end{abstract}


\section{Introduction}\label{20170822-sec1}

In a seminal paper \cite{Silling2000}, Silling introduced a nonlocal theory of elasticity in integral form that accounts for the interaction between continuum particles at finite distances, reminiscent of models involving interatomic interactions. The theory, called Peridynamics, has subsequently been extended to situations involving non-central forces and inelasticity; see for example the review article \cite{Silling2010}.
It has recently gained significant attention, at least in part because of the advantages it affords numerical calculations involving discontinuities\footnote{For example, Sun and Sundararaghavan \cite{Sun2014} show in their study of crystal plasticity that peridynamics can model finer shear bands with less simulation time than the traditional finite element solution.}.
Since the theory does not involve displacement gradients, it can handle geometric singularities with relative ease and is  therefore particularly well suited for the study of defects. This is the motivation for the present study where we analytically examine the propagation of a certain type of defect -- a kink -- according to peridynamics.

Analytical solutions within the peridynamic theory are few, and limited to linear problems, such as determining Green's functions, Weckner et al. \cite{Weckner2009}, Wang et al. \cite{Wang2017}, and using them to study the static and dynamic response of an infinite bar, Silling et al. \cite{Silling2003}, Weckner and Abeyaratne \cite{Weckner2005} and Mikata \cite{Mikata2012}.   While the peridynamic operator in our study is linear, the effective body force is nonlinear, because accounting for the defect requires the material to have two states, and therefore the model  to involve a double-well potential.  However, by taking this potential to be bi-quadratic, we are able to analytically solve the peridynamic defect propagation problem.

Energy functions with multiple local minima (energy-wells) are frequently encountered when modeling various physical systems.  In equilibrium, the system maybe ``stuck'' in a metastable energy-well, and  under a suitably large external stimulus, e.g. force, will undergo a progressive transition from the metastable state towards a stable state. Often, such an evolution involves a kink transition front that transforms the system locally as it passes through.
Such phenomena are observed in, for example, mechanical systems (friction \cite{Ward2015},  dynamics of CNT foams \cite{Thevamaran2014},   lattices of bistable buckled elastic structures \cite{Nadkarni2014}, and  mechanical transmission lines  \cite{Scott1969});   material systems (dislocation motion \cite{FK1938}, ferromagnetic domain wall motion \cite{Bishop1979},  commensurate phase transitions \cite{Braun1997}, and  chemical surface adsorption \cite{Rotermund1991}); electromagnetic systems (magnetic flux propagation in Josephson junctions  \cite{Stewart1968});   biological systems (pulse propagation in neurophysiology \cite{Scott1975} and rotation of DNA bases \cite{Peyrard1989});  and even traffic flow \cite{Braun1998}.  Determining the speed with which the system transitions into the stable state, as a function of the applied stimulus, is a question of significant interest.

There is a considerable literature on the propagation of kink transition fronts in lattices. The model proposed by Frenkel and Kontorova (FK) \cite{FK1938}, that now goes by their names, underlies many such studies. Atkinson and Cabrera \cite{AC1965} made an important contribution in the analysis of the FK model where, by using a bi-quadratic double well potential (rather than a trigonometric one) they solved the problem in analytic closed form. In particular, they derived a relation between applied force and kink propagation speed, and demonstrated the apparent dissipation of energy due to its radiation by waves propagating away from the kink. Several important contributions concerning kink propagation through a discrete lattice have been made by Truskinovsky, Vainchtein and their collaborators, e.g. \cite{Trusk1990, Vainchtein2013,Trusk2003, Trusk2005a, Trusk2005b, Trusk2006, Vainchtein2012}; for a  recent study, see Shiroky and Gendelman \cite{Shiroky2017}.  A survey of this literature can be found in the book by Braun and Kivshar \cite{Braun2004}; see also Slepyan  \cite{Slepyan-Book}.  A striking characteristic of Atkinson and Cabrera's solution  is that it predicts unbounded values of force at certain small values of speed.

Turning next to continuum models of kink propagation, the simplest is where the lattice is replaced by a linear elastic solid (but with a bi-quadratic double-well onsite potential). Here one finds that the force must equal the Maxwell force $f_0$ for {\it all} subsonic propagation speeds, and must take a different value $f_M (> f_0)$ for {\it all} supersonic speeds, e.g. see \cite{Vedantam1999, Kresse2002, Kresse2003}. This indeterminacy of the speed corresponding to a given value of force is likely due to the absence of a length scale in classical elasticity;  after all, the discreteness of a lattice model introduces a length scale into the problem even if only nearest neighbor interactions are accounted for. The next simplest continuum model is obtained by replacing the lattice by an elastic solid that includes strain gradient effects -- the so-called Boussinesq approximation. Kink propagation according to this model has been studied in \cite{Vedantam1999, Kresse2002, Kresse2003} but, as noted by Kresse and Truskinovsy \cite{Kresse2003}, is  a poor approximation at small wave lengths due to instability.  Kresse and Truskinovsy \cite{Kresse2003} and Truskinovsky and Vainchtein \cite{Trusk2006} have studied an alternative model that does not suffer from this deficiency, the so-called quasi-continuum model that is based on an approach put forward by Rosenau \cite{Rosenau1987}.  The quasi-continuum model may be viewed as arising from the retention of `micro-inertia'' terms in Mindlin's strain gradient theory \cite{Mindlin1964}.  More elaborate kink propagation models based on more detailed atomistic considerations have also been studied, e.g. Abeyaratne and Vedantam \cite{Vedantam2003} and  Hildebrand and Abeyaratne \cite{Hildebrand2008}.  Results pertaining to FK models in higher dimensions  are described in Chapter 11 of \cite{Braun2004}.

In the present paper a discrete lattice is approximated by a peridynamic model, and this is used to study kink propagation. The peridynamic operator is taken to be linear, bond-based and involves a single ``micro-modulus function''.  However the governing equation is nonlinear due to the presence of the double-well onsite potential that occurs in the body force term outside the peridynamic operator\footnote{{{Double-well potentials arise in modeling both kink and phase boundary propagation, the mathematical distinction being that in the latter case the nonlinear potential occurs within the peridynamic operator as in Dayal and Bhattacharya \cite{Dayal2006}. The analysis in this case is considerably more difficult, e.g. the solution in \cite{Dayal2006} is numerical in contrast to the analytical solution we find here.}}}.  Motivated by Atkinson and Cabrera \cite{AC1965}, the double well potential is taken to be bi-quadratic.  This introduces three parameters: a modulus $\mu$, distance between energy wells $d$, and the height of the energy barrier $\mu u_*^2/2$.  As for the micromodulus function, its simplest form involves two material parameters, an elastic modulus $E$ and a length scale $\ell$, whose values are chosen by matching features of the dispersion relations of the peridynamic and discrete lattice models. The particular form of the micromodulus function we use is motivated by Silling \cite{Silling2014}, and remarkably, is also implied by the quasi-continuum model as we shall show. Using Fourier transformation, we solve in closed form the static problem for an equilibrium kink and the dynamic problem for a steadily propagating kink. A kink will be in equilibrium if the applied force is smaller than a certain value\footnote{The fact that $\mathsf{F}_{\rm min}$ exceeds the Maxwell force $f_0$ is said to describe the phenomenon of ``lattice trapping''.} $\mathsf{F}_{\rm min} (> f_0)$. When the force exceeds this value, the kink can propagate at a steady speed given by a kinetic law that is determined by the analysis. This law relates the applied force to the propagation speed. As the speed increases, the force needed to propagate the kink increases monotonically from $\mathsf{F}_{\rm min}$ to $f_M$. The dynamic solution involves effective damping caused by energy radiation away from the kink and is well-behaved at all velocities.

After solving the kink propagation problem using peridynamic theory, we observed that the results had a similar form to those arising from the quasi-continuum model.  Indeed, we found that for a particular choice of the ratio of the peridynamic and lattice length scales, $\ell/d$, the results coincided exactly.  This led us in Section \ref{20170915-sec-last} to look at the relation between peridynamic theory and differential equation models of elastic continua more generally in settings not limited to kink propagation. In Section \ref{20170915-sec-last} we do {\it not} assume the form of the micromodulus function a priori, but rather determine it such that the results of the peridynamic theory and a class of differential-equation-based theory coincide.  Using an approach based on Whitham \cite{Whitham1999}, a formula for calculating the equivalent micromodulus function is derived. It is then possible to replace a differential-equation-based model with an equivalent peridynamic one, and this may prove useful when numerically studying more complicated problems such as those involving multiple and interacting defects.

The  paper is organized as follows: first, the basic one-dimensional discrete model of kink propagation through a lattice is described in Section \ref{20170822-sec0}.  For purposes of comparison, in Section \ref{20170912-sec1} we state (without proof) the results of the classical linear elastic solution of this problem. In Section \ref{20170826-sec-888} we describe the linear bond-based peridynamic model; show how its parameters can be chosen by comparison with the lattice model; and then go on the solve the static and dynamic problems for a kink.  In particular, the kinetic law relating the applied force to the kink speed is derived. In Section \ref{20170915-sec-last} we compare the results of the previous section with those of the quasi-continuum theory, and this leads us to explore the relation between the peridynamic model and other differential-equation-based continuum models in a more general dynamical setting.

%%%%%%%%%%%%%%%%%%%%%%%%%%%%%%%%%%%%%%%%%%%%%%%
%%%%%%%%%%%%%%%%%%%%%%%%%%%%%%%%%%%%%%%%%%%%%%%

\section{Preliminaries.}\label{20170822-sec0}

Consider a one-dimensional row of particles.   In a reference configuration, they are equally spaced, a distance $d$ apart, with the $n$th particle located at $x = nd$, $n=0, \pm 1, \pm 2, \ldots$.  During a motion, the displacement of the $n$th particle at time $t$  is $u_n(t)$.
%%%%%
\begin{figure}[h]
\centerline{\includegraphics[scale=0.65]{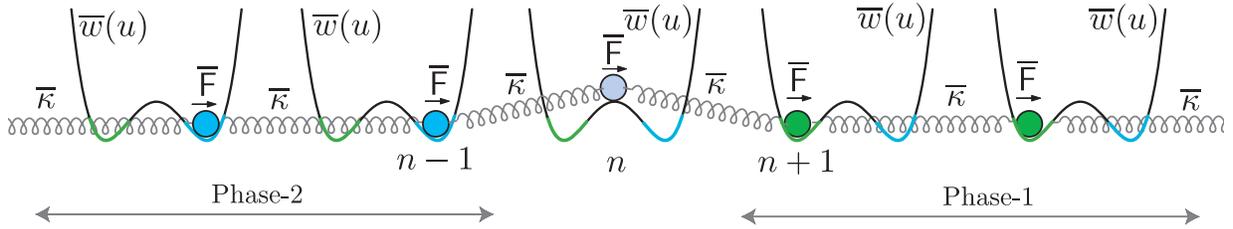}}
\caption{\footnotesize  Schematic depiction of a one-dimensional row of particles with nearest neighbor interactions. Each particle is also  associated with an onsite double-well potential $\overline{w}(u)$.}
\label{fig-a1}
\end{figure}
%%%%%

Each particle has mass $\mathsf{m}$ and is connected to
its $N$ nearest neighbors by  linear springs.  Specifically, for each $m = \pm 1, \pm 2, \ldots \pm N$, the $n$th particle is connected to the $n+m$th  particle  by a spring of stiffness $\overline{\kappa}_m$ and unstretched length $md$.
Each particle is also attached to a fixed foundation by a nonlinear spring characterized by a double-well potential $\overline{w}(u)$ as shown schematically in Figure \ref{fig-a1}. The local minima of $\overline{w}$ are at $u=0$ and $u=d$.  The effect of the external loading  is accounted for by a force
$\overline{\mathsf{F}}$ acting on each particle.
The equation of motion of the $nth$ particle is therefore
\be\label{a1}
\sum_{m=1}^N \overline{\kappa}_m \big[ u_{n+m}(t) - 2 u_n(t) + u_{n-m}(t) \big] +  \overline{\mathsf{F}}  - \overline{w}'\big(u_n(t)\big) = \mathsf{m} \ddot{u}_n(t)  , \quad n = 0, \pm 1,  \ldots.
\ee
It is convenient to write
$$
u_n(t) = u(x,t), \qquad x = n d,
$$
where $u(x,t)$ is a suitably smooth function defined for $- \infty < x < \infty, - \infty < t < \infty$.
Then \eqref{a1} can be written as the delay-differential equation
\be\label{20170826-gerol-eq5}
\sum_{m=1}^N \kappa_m  \, \frac{u(x+md,t) - 2 u(x,t) + u(x-md, t)}{d^2} +  \mathsf{F} - w'\big(u(x, t)\big) = \rho {u}_{tt}(x,t) ,
\ee
where we have let $\rho = \mathsf{m}/d^3, \kappa_m = \overline{\kappa}_m/d, {\mathsf{F}} = \overline{\mathsf{F}}/d^3$ and $w = \overline{w}/d^3$.

In order to obtain explicit closed-form solutions to the problem to be studied, we take the  double-well potential $w$ to be bi-quadratic:
\be\label{a15}
w(u) =  \left\{
\ba{llll}
\displaystyle \frac{{\footnotesize{\mathsf{\mu}}}}{2} u^2, &\qquad u \leq u_*,\\[2ex]

\displaystyle \frac{{\footnotesize{\mu}}}{2} (u-d)^2 +\mu\left( u_* - { \footnotesize \frac 12} d\right)d, &\qquad u \geq u_*,\\
\ea
\right.
\ee
where $\mu >0$ and $u_*>0$ are constant parameters. Observe that $w$ has local minima -- energy-wells -- at $u=0$ and $u=d$, and a local maximum at $u=u_*$.  If the displacement of some particle takes a value $< u_*$ we say it is in ``phase-1''; if  it exceeds $u_*$, we say it is in ``phase-2''. The force\footnote{For simplicity of terminology we refer to ${\mathsf{F}}, f=w', f_m, f_M$ and $f_0$ as ``forces'' though they are in fact force densities: forces per unit volume.
} $f = f(u) = w'(u)$ associated with this potential is bi-linear:
\be\label{gerol-eq2}
f(u) = w'(u) =\left\{
\ba{llll}
\displaystyle {{\footnotesize{\mathsf{\mu}}}} u, &\qquad u < u_*,\\[2ex]

\begin{figure}[h]
\centerline{\includegraphics[scale=0.65]{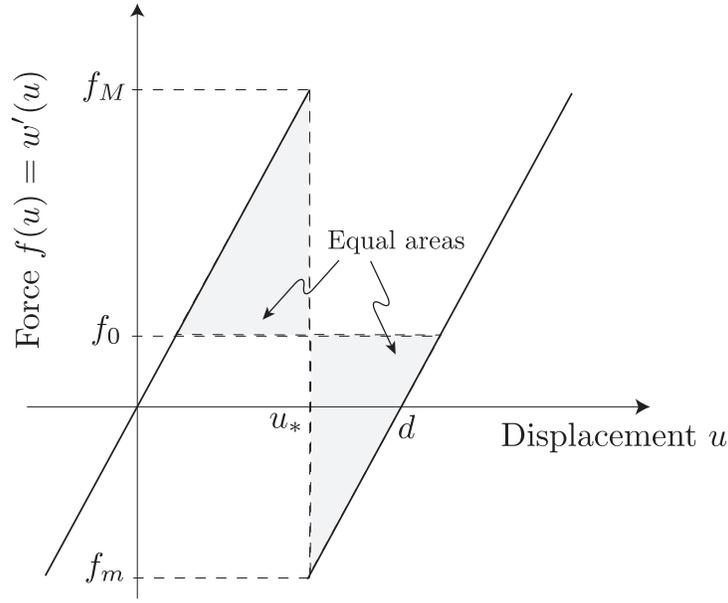}}
\caption{\footnotesize  Force $f$ versus displacement $u$  according to \eqref{gerol-eq2}.}
\label{20170708-fig-ss}
\end{figure}

\displaystyle {{\footnotesize{\mu}}} (u-d), &\qquad u > u_*.\\
\ea
\right.
\ee
It is convenient to let $f_m, f_M$ and $f_0$ denote the force levels
\be\label{20170716-eq1}
f_m \coloneqq f(u_*^+) = {{\footnotesize{\mu}}} (u_*-d), \qquad f_M \coloneqq f(u_*^-)= \mu u_*, \qquad f_0 \coloneqq \frac12 (f_m + f_M) =
{{\footnotesize{\mu}}} \left(u_*- \frac{d}{2}\right),
\ee
whose meanings are shown in Figure \ref{20170708-fig-ss}.  A particle cannot be in phase-1 if the force exceeds $f_M$, and it cannot be in phase-2 if the force is smaller than $f_m$.  The {\it Maxwell force-level} $f_0$ cuts off lobes of equal area as shown in the figure. It is the force at which the two phases have the same potential energy.

In the particular case of a {\it uniform equilibrium configuration} in which all particles have the same displacement $u(x,t) = u$, it follows from \eqref{20170826-gerol-eq5}  that $u$ is given by the roots of the equation $w'(u) = \mathsf{F}$.   For the bi-quadratic energy \eqref{a15}, when the applied force $\mathsf{F}$ lies in the intermediate range
\be\label{e50}
f_m < \mathsf{F} < f_M,
\ee
(see Figure \ref{20170708-fig-ss}),  this leads to two possible equilibrium displacements $u^\pm$:
\be\label{a13}
{u}^{+} \coloneqq \frac{{\mathsf{F}}}{{\footnotesize{\mu}}}, \qquad {u}^{-} \coloneqq
\frac{{\mathsf{F}}}{{\footnotesize{\mu}}} + d.
\ee
The displacement $u^+$ lies in phase-1, $u^-$ in phase-2. We assume throughout that the value of the force $\mathsf{F}$ is given, that it lies in the range \eqref{e50}, and that
 $u^\pm$ are then defined by \eqref{a13}. They necessarily obey
$$
u^+ < u_* < u^-.
$$

In Sections \ref{20170912-sec1} and \ref{20170826-sec-888}  our primary interest will be in the steady motion of a ``kink''  associated with the motion of particles from phase-1  to phase-2, {\it one at a time}.  At a typical instant, all particles ahead of the $n$th one are still in phase-1 while those behind it have moved into phase-2. Suppose that
during this motion the
$nth$ particle stays in phase-1 for times $t < n\tau$ (for some $\tau$);
at time $t=n\tau$ it moves into
phase-2 and remains there for $t > n\tau$:
$$
u_n(t) \  \left\{
\ba{lll}
< u_*, \qquad \qquad &t < n \tau,\\[2ex]

= u_*, \qquad \qquad &t = n \tau,\\[2ex]

> u_*, \qquad \qquad &t > n \tau.\\
\ea \right.
$$
 This motion continues steadily with the $(n+1)$th particle
moving into phase-2 at time $t=(n+1)\tau$ and so on.  Such a steady motion is described by
\be\label{a10}
u_{n}(t ) = u_{n+m}(t + m\tau)\qquad {\rm for \ all \ integers} \  n,m.
\ee
Setting $m=-n$ in \eqref{a10} shows that $u_{n}(t ) = u_{0}(t - n\tau) = u_{0}(t - nd/c)$ where $c ={d}/{\tau}$ is the propagation speed of  the kink.
Therefore the displacement
field of interest has the traveling wave form
$$
u_n(t) = u(z) \qquad {\rm where} \ z = nd - ct.
$$
The particles far ahead of the kink are taken to be in equibrium is phase-1 so that $u_n(t) \to u^+$ as $n \to \infty$, i.e. $u(z) \to u^+$ as $z \to \infty$.

We use the term ``kink'' to refer to a configuration, either static or dynamic, in which all particles ahead of a particular one are in phase-1 and those behind it are in phase-2 .

%%%%%%%%%%%%%%%%%%%%%%%%%%%%%%%%%%%
%%%%%%%%%%%%%%%%%%%%%%%%%%%%%%%%%%%

\section{\bf Classical elasticity model.} \label{20170912-sec1}

The simpest mathematical model of the problem described above is obtained by replacing the finite difference terms in \eqref{20170826-gerol-eq5} by the second spatial derivative of $u$ leading to
$$
E u_{xx} + \mathsf{F} - w'(u) = \rho u_{tt}, \qquad E = \sum_{m=1}^N m^2 \kappa_m.
$$
The associated material wave speed is  $c_0 = \sqrt{E/\rho}.$
Steady traveling wave solutions, $u=u(z), z = x - ct$, of this equation have been studied, e.g. see \cite{Vedantam1999, Kresse2002},  with particles ahead of $z=0$ being in phase-1, those behind it in phase-2, and
$$
u(z) \to u^+ \qquad {\rm as } \ z \to \infty.
$$
The displacement and strain, $u(z)$ and $u'(z)$, are required to be continuous. In the case of subsonic propagation $c < c_0$ the solution is
\be\label{20170912-eq1}
u(z) = \left\{
\ba{lll}
\displaystyle u^+ + \frac{u^- - u^+}{2} \e^{- \beta z}, \qquad &z \geq 0,\\[2ex]

\displaystyle u^- - \frac{u^- - u^+}{2} \e^{ \beta z}, \qquad &z \leq 0,\\
\ea
\right. \qquad {\rm where} \quad  \beta = \sqrt{\frac{\mu}{(c_0^2 - c^2)}} > 0,
\ee
together with
\be\label{20170912-eq2}
\mathsf{F} = f_0, \qquad 0 < c < c_0,
\ee
where \eqref{20170912-eq2} follows from the requirement $u(0)=u_*$ necessitated by the continuity of $u$. Thus in the subsonic case, no matter what the propagation speed, the force must equal the Maxwell force $f_0$.

On the other hand for supersonic propagation speeds $c > c_0$ one finds
\be\label{20170912-eq3}
u(z) = \left\{
\ba{lll}
u^+ , \qquad &z \geq 0,\\[2ex]

u^- - (u^- - u^+) \cos \alpha z  , \qquad &z \leq 0,\\
\ea
\right. \qquad {\rm where} \quad  \alpha = \sqrt{ \frac{\mu}{(c^2 - c_0^2)}} > 0,
\ee
with
\be\label{20170912-eq4}
\mathsf{F} = f_M, \qquad c > c_0.
\ee
Thus in the supersonic case, no matter what the propagation speed, the force must equal the maximum force level $f_M$  indicated in Figure \ref{20170708-fig-ss}.

\begin{figure}[h]
\centerline{\includegraphics[scale=0.3]{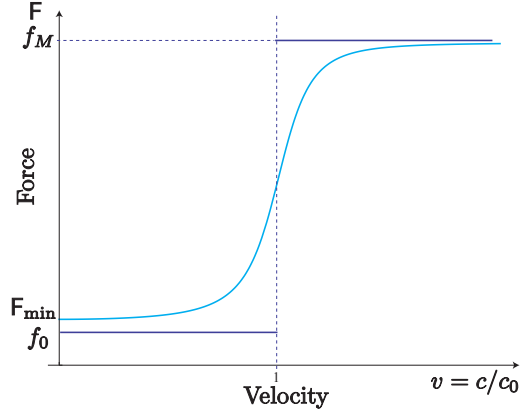}}
\caption{\footnotesize Kinetic relation: the two solid horizontal  lines correspond to the relation between $\mathsf{F}$ and $v=c/c_0$ according to classical elasticity \eqref{20170912-eq2}, \eqref{20170912-eq4}.  The curve corresponds to the kinetic relation \eqref{PD-D21} determined later according to the peridynamic model.}
\label{Fig-SingleKinReln2}
\end{figure}

The horizontal solid lines in Figure \ref{Fig-SingleKinReln2} show the relation between the force $\mathsf{F}$ and propagation speed $c/c_0$ as described by \eqref{20170912-eq2} and \eqref{20170912-eq4}. (The curve corresponds to the peridynamic solution to be determined later.) Observe how  in the subsonic case  the solution \eqref{20170912-eq1} decays exponentially away from the kink in both directions, but  in the supersonic case involves oscillatory waves that are radiated without decay behind the kink, see \eqref{20170912-eq3}. The radiated energy in these waves describes ``dissipation''; this is also implied by the fact that $\mathsf{F}$ {\it exceeds} the Maxwell-force in this case.

%%%%%%%%%%%%%%%%%%%%%%%%%%%%%%%%%%%
%%%%%%%%%%%%%%%%%%%%%%%%%%%%%%%%%%%

\section{Peridynamic model.}\label{20170826-sec-888}

A {\it peridynamic model} is a nonlocal continuum model that accounts for the effect of a neighborhood on a material point.  In (linear, homogeneous, bond-based) peridynamics, the force per unit length applied on the particle at $x$ by the particle at $x+\xi$ is
$$
C(\xi) \big[u(x+\xi, t) - u(x, t)\big],
$$
and so the equation of motion \eqref{20170826-gerol-eq5} is replaced by its peridynamic counterpart
\be\label{gerol-eq13}
\int_{-\infty}^\infty C(\xi) \big[ u(x+\xi,t) - u(x,t)\big]\, d\xi \  + \  \mathsf{F} - w'\big(u(x, t)\big) = \rho {u}_{tt}.
\ee
 The {\it micromodulus function} $C(\xi)$ in \eqref{gerol-eq13} characterizes the material.  On physical grounds $C$ is  required  $(i)$ to be symmetric: $C(\xi) = - C(-\xi)$; $(ii)$ to decay at infinity: $C(\xi) \to 0$ as $\xi \to \pm \infty$; and $(iii)$ to involve a length scale $\ell$ such that $C(\xi)$ approaches a suitable generalized function in the limit $\ell \to 0$.

Note that since particles interact at a distance, it is not necessary that the displacement field in peridynamic theory be continuous. For example suppose that the micromodulus function $C(x)$ is nonzero for $|x| < h$ and vanishes identically for $|x| > h$ for some $h >0$ (the ``horizon''). The continuum would then remain coherent even in the presence of displacement discontinuities of magnitude less than $h$.

Throughout {\it this section} we shall consider a micromodulus function of the form
\be\label{gerol-eq14}
C(x) =  \frac{E}{2 \ell^3}
\, \e^{-|x|/\ell},
\ee
where the longitudinal stiffness $E>0$ and the length $\ell > 0$ are material constants. This form is motivated by an analysis of nonlocal effects by Silling \cite{Silling2014}.
It will be convenient in what follows to let {Deleted $\alpha$ from next equation.}
\be\label{gerol-eq20}
\beta \ \coloneqq \  \frac{\mu \ell^2}{E},
\ee
and the intrinsic {\it wave speed} $c_0$ associated with this material model by
\be\label{31421}
c_0 \coloneqq  \sqrt{{E}/{\rho}}.
\ee

Observe that if instead of \eqref{gerol-eq14}, the micromodulus function is taken to be
\be\label{20170826-eq1}
C_D(x) = \sum_{m=1}^N \frac{\kappa_m}{d^2} \big[ \delta(x - md) + \delta(x+md)\big],
\ee
where $\delta$ denotes the Dirac $\delta$-function, the peridynamic equation of motion \eqref{gerol-eq13} coincides with the equation of motion \eqref{20170826-gerol-eq5} of the discrete model.
If we want \eqref{gerol-eq14} to approximate  \eqref{20170826-eq1}, we can choose the material parameters $E$ and $\ell$ in \eqref{gerol-eq14} by matching, for example, the dispersion relations of the two models.
By looking at motions of the form $u(x,t) = \e^{i(kx-\omega t)}$ of equations \eqref{20170826-gerol-eq5} and \eqref{gerol-eq13} (with $\mathsf{F}$ and $w'(u)$ set equal to zero) one obtains the following respective dispersion relations relating the frequency $\omega$ to the wave number $k$:
\be \label{20170913-eq10}
\left.
\ba{llll}
{\rm Discrete} \qquad &\rho \omega^2 &= \displaystyle \int_{-\infty}^\infty C_D(\xi) \big( 1 - \cos k\xi \big) \, d\xi  \ &\displaystyle = \ \sum_{m=1}^N 2\frac{\kappa_m}{d^2} \big( 1 - \cos m kd\big), \\[3ex]

{\rm Peridynamic} \qquad & \rho \omega^2 &= \displaystyle \int_{-\infty}^\infty C(\xi) \big( 1 - \cos k\xi \big) \, d\xi  &\displaystyle = \frac{E}{d^2} \, \frac{(kd)^2}{1 + (kd)^2 \, \ell^2/d^2}.\\
\ea\right\}
\ee

\begin{figure}[h]
\centerline{\includegraphics[scale=0.35]{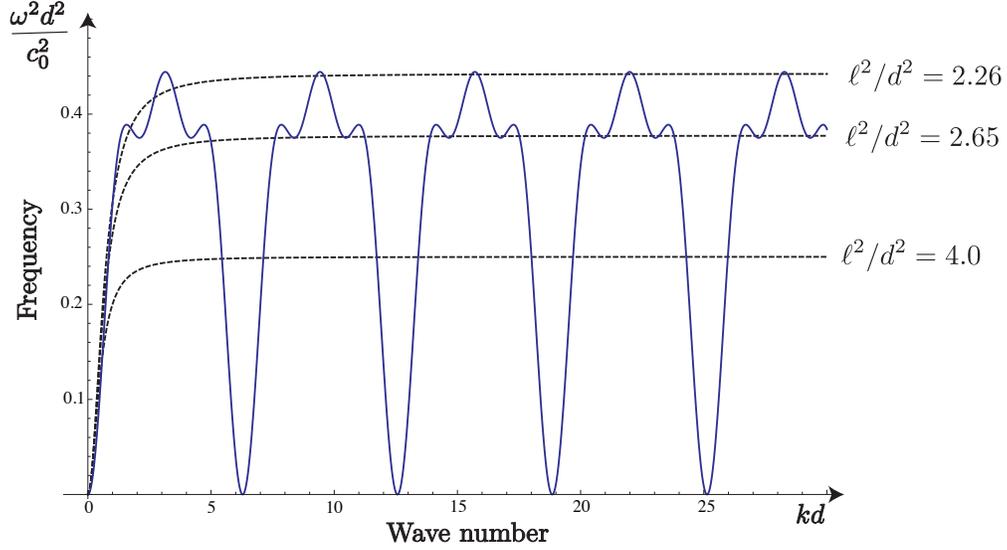}}
\caption{\footnotesize  Dispersion relation of the discrete model with $N=3$ (solid); and that of the peridynamic model for different values of $\ell/d$ (dashed).}
\label{20170708-fig-dispersionreln-new}
\end{figure}

The parameters $E$ and $\ell$ can be determined by matching certain features of these two dispersion relations. For example,
equating \eqref{20170913-eq10}$_1$ and  \eqref{20170913-eq10}$_2$ in the long wavelength limit, $kd \to 0$,  yields
$$
E =    \sum_{m=1}^N m^2\kappa_m.
$$
In the shortwave length limit, $kd \to \infty$,  the peridynamic dispersion relation \eqref{20170913-eq10}$_2$ gives
$$
\frac{\omega^2d^2}{c_0^2} \to \frac{d^2}{\ell^2}.
$$
The discrete dispersion relation oscillates in this limit and so the value of $\ell/d$ can be chosen by matching some desirable average feature. Figure \ref{20170708-fig-dispersionreln-new} shows plots of the dispersion relations  \eqref{20170913-eq10}$_1$ and  \eqref{20170913-eq10}$_2$. In drawing the figures we have taken $\kappa_m = \kappa/m$, $N=3$, $\displaystyle E = \kappa N(N+1)$ and plotted $\omega^2 d^2/c_0^2$ versus $kd$. Three peridynamic dispersion relations are shown corresponding to three values of $\ell/d$ illustrating how one can vary the way in the peridynamic and discrete dispersion relations match by varying this parameter. One could of course work with a micromodulus function with additional parameters and match additional details of the dispersion relations.

%%

%%%%%%%%%%%%%%%%%%%

Finally, it is useful to introduce the function $M(k)$, defined in terms of the micromodulus function by
\be\label{20170715-eq10}
M(k) \coloneqq \int_{-\infty}^{\infty} C(\xi) \big[ 1 - \e^{-ik\xi}\big] \, d\xi = \int_{-\infty}^{\infty} C(\xi) \big[ 1 - \e^{ik\xi}\big] \, d\xi= \int_{-\infty}^{\infty} C(\xi) \big[ 1 - \cos k \xi \big] \, d\xi,
\ee
where the second and third equalities follow since $C(\xi)$ is an even function.  $M$ will play a central role in what follows (and already appeared in \eqref{20170913-eq10}$_2$ above). It is a one-dimensional scalar counterpart of the acoustic tensor, see \cite{Weckner2005}, and so we will refer to it as the {\it acoustic function}.
When the micromodulus function has the exponential form \eqref{gerol-eq14}, the acoustic function specializes to
$$
 M(k) =  {E} \, \frac{k^2}{1+ k^2\ell^2} .
$$

In summary, the peridynamic model considered in this section is characterized by \eqref{gerol-eq13} and
\eqref{gerol-eq14}.

%%%%%%%%%%%%%%%%%%%%%%%%%

\subsection{Equilibrium configuration of a kink.} \label{20170823-sec-statics}

For an equilibrium configuration $u(x,t) = u(x)$, the peridynamic equation of motion \eqref{gerol-eq13} specializes to the equilibrium equation
\be\label{ep2}
\int_{-\infty}^\infty C(\xi) \big[ u(x+\xi) - u(x)\big]\, d\xi \  + \  \mathsf{F}  - w'(u(x)) = 0.
\ee
In the particular case of a uniform equilibrium configuration, $u(x) =u$ for all $x$, with the force lying in the range \eqref{e50}, equations \eqref{ep2} and \eqref{gerol-eq2} tell us that the displacement must take one of the values $u^+ = \mathsf{F}/\mu$ or  $u^- = \mathsf{F}/\mu + d$.

We now consider {\it non-uniform} equilibrium configurations in which
 all particles $x>0$ are associated with phase-1, and  all particles $x<0$ are associated with phase-2,
with
\be\label{eqc2}
u(x) \to u^\pm \quad {\rm as} \ x \to \pm \infty.
\ee
For the class of displacement fields $u(x)$ under consideration,
\be\label{eqc1}
u(x) \ \left\{
\ba{lll}
< u_* \qquad {\rm for} \ x>0,\\[2ex]

> u_* \qquad {\rm for} \ x<0.\\
\ea
\right.
\ee
Observe from \eqref{ep2} that because of the discontinuity of $w'(u)$,  $u(x)$ must be discontinuous at $x=0$ and so we are {\it not} requiring $u(0) = u_*$ in \eqref{eqc1}.
Writing \eqref{eqc1} as $H(u(x) - u_*) = H(-x)$ where $H(x)$ is the Heaviside step function, and using this in the bi-quadratic energy function \eqref{a15} leads to
$$
w'(u(x)) \, = \, \mu u(x) - \mu d \, H(-x).
$$
Therefore the equilibrium equation  \eqref{ep2} specializes to
\be\label{ep3}
\int_{-\infty}^\infty C(\xi) \big[ u(x+\xi) - u(x)\big]\, d\xi \   - \mu u(x)  \ = \  - \mu u^- - \mu(u^+ - u^-) H(x).
\ee
Given a force $\mathsf{F}$ whose value lies in the range \eqref{e50}, an equilibrium displacement field according to the peridynamic model is a function $u(x)$ that satisfies
\eqref{ep3}, \eqref{eqc2} and \eqref{eqc1}, with $u^\pm$ given by \eqref{a13}.

As observed already, the discontinuity of $w'(u(x))$ at $x=0$ requires $u(x)$ to be discontinuous at $x=0$. To calculate the jump in the value of $u$ at $x=0$,  we evaluate \eqref{ep3} at $x= +\varepsilon >0$ and at $x = -\varepsilon <0$ and subtract one equation from the other. Letting $\varepsilon \to 0$ in the result leads to
\be\label{20170718-eq1}
-\ljump u \rjump \, \int_{-\infty}^\infty C(\xi) \, d\xi \   - \mu \ljump u \rjump  = - \mu (u^+ - u^-),
\ee
where we have used the standard notation $\ljump g \rjump$ for the jump in the value of a function $g(x)$ at $x=0$:
$$
\ljump g \rjump \ \coloneqq \ \lim_{\varepsilon \to 0} \left\{ g(+\varepsilon) - g(-\varepsilon) \right\}.
$$
Thus the displacement $u$ suffers a jump discontinuity at $x=0$ of magnitude
\be\label{20170718-eq3}
\ljump u \rjump = - \, \frac{\mu d}{\mu + \int_{-\infty}^\infty C(\xi) \, d\xi },
\ee
where we have used \eqref{a13}.
For the exponential micromodulus function \eqref{gerol-eq14} this specializes to
\be\label{20170718-eq4}
\ljump u \rjump = - d \, \frac{\beta}{ 1+ \beta},
\ee
where $\beta$ was introduced in \eqref{gerol-eq20}. { Displacement jumps in the peridynamic theory were encountered in \cite{Weckner2005}, where it was shown in particular, that such discontinuities cannot propagate. Indeed, when we consider the propagating kink in the next section, we find no displacement discontinuities.}

The solution of \eqref{ep3}, \eqref{eqc2}, \eqref{eqc1} can be readily determined by Fourier transforming \eqref{ep3} with \eqref{gerol-eq14} in mind. This leads to
\be\label{20180823-eq1}
u(x)=\frac{u^{+}+u^{-}}{2} + \frac{\mu d}{2\pi{\rm{i}}}\int_{-\infty}^{+\infty}\frac{e^{-{\rm{i}}kx}}{k L(k)} \de k,
\ee
where
\be\label{20160928-eq9}
L(k) = M(k) + \mu = \frac{\mu d^2}{b^2(k^2\ell^2 +1)}\left( k^2 + \frac{b^2}{d^2}\right),
\ee
and we have  set
\be\label{20160827-eq11}
b \coloneqq \frac{d}{\ell} \, \frac{1}{\sqrt{1 + 1/\beta}}.
\ee
The integral in \eqref{20180823-eq1}, interpreted as its Cauchy principal value, can be evaluated by contour integration using the Residue Theorem.

To do this, note first by \eqref{20160928-eq9}, that the integrand in \eqref{20180823-eq1} has three poles, one corresponding to $k=0$ and the other two being the two (purely imaginary) zeros of $L(k)$. Next,
 one considers a path $\Gamma$ along the real axis of the complex $k$-plane, indented by a small semi-circle $\calS_\varepsilon$ in the upper half plane\footnote{Indenting in the upper half-plane rather than the lower half plane is necessary in order to get the correct conditions at infinity.} of radius $\varepsilon$ centered at $k=0${; the complex $k$-plane and integration path are shown in Figure \ref{fig-Fig-ComplexPlane1} of the Supplementary Material}. The desired integral is then the limit as $\varepsilon \to 0$ of the difference between the integral on $\Gamma$ minus the integral on $\calS_\varepsilon$. The integral on $\calS_\varepsilon$, in the limit $\varepsilon \to 0$, is readily shown to be $- d/2$.  Thus \eqref{20180823-eq1} can be written as
$$
u(x) = u^- + \frac{\mu d}{2\pi i} \int_{\Gamma} \frac{\e^{-ikx}}{kL(k)} \, dk .
$$
This integral on $\Gamma$ can now be evaluated by completing the integration path by a large semi-circle in the lower half-plane for $x>0$ and the upper half plane for $x<0$ and using the Residue Theorem.
This leads to
\be\label{20160823-eq2}
u(x) = \left\{
\ba{lll}
       \displaystyle
       u^++ \frac d2 \, \frac{1}{1 + \beta} \, e^{-b x/d},  &  x>0, \\[2ex]
        \displaystyle
       u^- - \frac d2 \, \frac{1}{1 + \beta} \,e^{b x/d},  &  x<0,
   \ea\right.
   \ee
with $\beta$ and $b$ given by \eqref{gerol-eq20} and \eqref{20160827-eq11} respectively.

Observe from \eqref{20160823-eq2} that
\be\label{20160823-eq3}
u(0^+) = u^++ \frac d2 \, \frac{1}{1 + \beta}  , \qquad u(0^-) = u^- - \frac d2 \, \frac{1}{1 + \beta} \ ,
\ee
which, since $u^- - u^+ = d$ by \eqref{a13}, confirms that the jump condition \eqref{20170718-eq4} holds.

Finally, for the solution \eqref{20160823-eq2} to be acceptable, $u(x)$ must lie in the ranges given in \eqref{eqc1}. This requires that $u(0^+) < u_* < u(0^-)$. In view of \eqref{20160823-eq3} and \eqref{a13} this necessitates the force $\mathsf{F}$ to lie in the range
\be\label{20170823-e9}
f_0 - \frac{\mu d}{2}\, \frac{1}{1+ 1/\beta} < {\mathsf{F}} < f_0 + \frac{\mu d}{2}\, \frac{1}{1+ 1/\beta},
\ee
where $f_0$ is the Maxwell force given in \eqref{20170716-eq1}$_3$.
It is not difficult to show using  \eqref{20170716-eq1} that the rightmost
expression in \eqref{20170823-e9} is $< f_M$ while the leftmost expression is $> f_m$. Thus the interval of force demarcated by \eqref{20170823-e9} is a subset of the interval $(f_m , f_M)$ that includes the Maxwell force $f_0$ in its interior.

%%%%%%%%%%%%%%%%%%%%%%%%%%%%%%%%%%%%%%

\subsection{Dynamics of a steadily propagating kink.}\label{20170823-sec-dynamics}

We now consider the steady motion of a kink propagating at some (to-be-determined) speed $c$. The displacement field in such a motion has the form
$$
u(x,t) = u(z) \qquad {\rm where} \ z = x-ct.
$$
Substituting this into the peridynamic equation of motion \eqref{gerol-eq13} gives
\be\label{20170824-eq13}
\rho  c^2 u''(z) - \int_{-\infty}^\infty C(\xi) \big[ u(z+\xi) - u(z)\big]\, d\xi \   + w'\big(u(z)\big) =  \mathsf{F}.
\ee
The material ahead of the propagating kink is in phase-1, that behind it is in phase-2:
\be\label{20170823-eqc1}
u(z) \ \left\{
\ba{lll}
< u_* \qquad {\rm for} \ z>0,\\[2ex]

= u_* \qquad {\rm for} \ z=0,\\[2ex]

> u_* \qquad {\rm for} \ z<0.\\
\ea
\right.
\ee
According to \eqref{20170824-eq13}, in the dynamic case the discontinuity in $w'(u(z))$ at $z=0$ requires that $u''(z)$ be discontinuous at $z=0$ but $u$ and $u'$ are permitted to be continuous.  Therefore, in particular, we have required $u(0) = u_*$ in \eqref{20170823-eqc1}. This is in contrast to the equilibrium problem.
As for the jump in $u''$, it is seen from \eqref{20170824-eq13} that $\rho  c^2\ljump u''\rjump + \ljump w'(u)\rjump = 0$ whence
\be\label{20170928-eq10}
\ljump u''\rjump = - \frac{\ljump w'(u)\rjump }{\rho  c^2} = - \frac{\mu d}{\rho  c^2}.
\ee

Finally, as for the far-field conditions we assume that the particles far ahead of the kink are in {\it equilibrium} in
phase-1. Thus, recalling \eqref{a13}$_1$, we take
\be\label{a14}
u(z) \to u^+ = \frac{\mathsf{F}}{\mu} \qquad {\rm as} \quad z
\rightarrow +
\infty.
\ee
As for the particles far behind the wavefront, we simply require them to be in phase-2 but {\it do not assume them to be in equilibrium}. Thus as $z \to - \infty$ we do not require anything beyond \eqref{20170823-eqc1}$_3$. This allows for particles to be oscillating in the phase-2 energy-well. We could impose the stronger requirement that the {\it average} displacement of the particles behind the wave be $u^-$ (where $u^-$ is given by \eqref{a13}) but it is not necessary that we do so. The results show that this comes out automatically.

On substituting the bi-quadratic form \eqref{a15} of the energy into \eqref{20170824-eq13} and keeping \eqref{20170823-eqc1} in mind, leads to
\be\label{20170823-eq13}
\ba{lll}
\rho  c^2 u''(z) - \int_{-\infty}^\infty C(\xi) \big[ u(z+\xi) &-&u(z) \big]\, d\xi \   +  \mu u(z) = \mu u^+ + \mu (u^+ - u^-) H(z).
\ea
\ee
 The solution of \eqref{20170823-eq13}, \eqref{20170823-eqc1}, \eqref{a14} can again be written down by Fourier transformation, leading to
 \be\label{PD-D3}
u(z)=\frac{u^{+}+u^{-}}{2} + \frac{\mu d}{2\pi{\rm{i}}}\int_{-\infty}^{+\infty}\frac{e^{-{\rm{i}}kz}}{kL(k)} \de k,
\ee
where, now,
 \be\label{PD-D4}
 \begin{split}
L(k)= M(k) + \mu - \rho  c^2k^2
&=-\frac{\mu d^4}{r^2b^2\left( k^2\ell^2+1\right)}\left(k^2-\frac{r^2}{d^2}\right)\left(k^2+ \frac{b^2}{d^2}\right),
\end{split}
\ee
with $r >0$ and $b>0$ being the (positive real) quantities
\be\label{PD-D20}
 \begin{split}
&r=\frac{d}{\ell\sqrt{2}} \, \frac{1}{v} \, \left[ \left(1 - v^2 + \beta   \right) +\sqrt{\left( 1 - v^2 + \beta   \right)^2+ 4 \beta  \, v^2}\right]^{1/2},\\
&b=\frac{d}{\ell\sqrt{2}}\, \frac{1}{v} \, \left[-\left(1- v^2 + \beta   \right)+\sqrt{\left( 1 - v^2 + \beta   \right)^2+  4 \beta   \, v^2}\right]^{1/2}.
\end{split}
\ee
Here we have let
\be\label{20170707-eq1}
v \ \coloneqq \  \frac{c}{c_0}
\ee
be the (as-yet unknown) nondimensional kink propagation speed, and used $\beta$ and $c_0$ defined previously in \eqref{gerol-eq20} and \eqref{31421}.
Substituting (\ref{PD-D4}) into (\ref{PD-D3}) yields
 \be\label{PD-D6}
u(z)=\frac{u^{+}+u^{-}}{2}-\frac{r^2b^2}{d^3} \frac{1}{2\pi i} \int_{-\infty}^{+\infty} \frac{1+k^2\ell^2}{k(k^2-r^2/d^2)(k^2+b^2/d^2)} \, {e^{-{\rm{i}}kz}}  \, \de k,
\ee
where the integral is interpreted in the sense of its Cauchy principal value.

The integral in \eqref{PD-D6} can again be evaluated by contour integration using the Residue Theorem in a manner similar to that used in the study of the equilibrium kink in Section \ref{20170823-sec-statics}. The main difference is that here,  the integrand has two poles $k=\pm r/d$ on the real $k$-axis (in addition to the pair of purely imaginary poles). When constructing the integration path $\Gamma$ we must indent it so that $\Gamma$ passes below these two poles. { The complex $k$-plane and integration path are shown in Figure \ref{fig-Fig-ComplexPlane2} of the Supplementary Material}.  The reason it must pass below (rather than above) these poles is because these poles lead to oscillatory terms $\cos rz/d$ and  $\sin rz/d$ in the solution, which cannot exist as $z \to +\infty$ since those particles are assumed to be in equilibrium.  On the other hand we are allowing the particles to vibrate as $z \to -\infty$ and so such oscillatory terms are admissible for $z<0$. By closing the path of integration as in Section \ref{20170823-sec-statics} and using the Residue Theorem we find
 \be\label{20170823-PD-D18}
u(z)=\begin{cases}
       \displaystyle
      u^+ +  \, \frac d2 \, \frac{1}{b^2+r^2} \, \left(r^2 - \frac{\beta d^2}{v^2 \ell^2} \right) \, e^{-bz/d}, &  z\geq 0, \\[2ex]

        \displaystyle
       u^- -  \,\frac d2 \, \frac{1}{b^2+r^2} \, \left(r^2 - \frac{\beta d^2}{v^2 \ell^2} \right) \, e^{bz/d} \, -  \, \frac{d}{b^2+r^2} \, \left( b^2 + \frac{\beta d^2}{v^2 \ell^2} \right) \, \cos\left(\frac{rz}{d}\right) , &  z\leq 0.
        \end{cases}
\ee
It can be readily verified that $u,u'$ and $u'''$ are continuous at $z=0$ and that $u''$ has the appropriate discontinuity \eqref{20170928-eq10}. The requirement $u(0)=u_*$ is yet to be enforced.

An alternative illuminating form of the solution is
\be\label{20170926-eq1}
u(z) = \left\{
\ba{lll}
\displaystyle u^+ + (u_* - u^+) \e^{-bz/d}, & z \geq 0,\\[2ex]
\displaystyle u^- - (u_* - u^+) \e^{bz/d} - (u^+ + u^- - 2 u_*) \, \cos\left(\frac{rz}{d}\right) , &  z\leq 0.\\
\ea\right.
\ee
Observe using \eqref{20170716-eq1} and \eqref{a13} that the coefficient of the exponential terms vanishes if $\mathsf{F} = f_M$ while the coefficient of the cosine term vanishes if $\mathsf{F} = f_0$.  The displacement field written in the form \eqref{20170926-eq1} satisfies the requirements $\ljump u \rjump = 0, \ljump u' \rjump = 0, \ljump u''' \rjump = 0$ and $u(0)=u_*$.
The requirement on $\ljump u'' \rjump$ in \eqref{20170928-eq10} remains to be enforced.

%%%%%%%%%%%%%%%%%%%%%

\subsection{Kinetic relation.}

While $u(z)$ given by \eqref{20170823-PD-D18} is continuous at $z=0$ we are yet to enforce the requirement $u(0)=u^*$ necessitated by \eqref{20170823-eqc1}$_2$. When this is enforced \eqref{20170823-PD-D18} leads to the following  relation between the applied force $\mathsf{F}$ and the kink propagation speed $v$:
 \be\label{PD-D19}
\mathsf{F}=f_0+\frac{\mu d}{2(b^2+r^2)}\left(b^2+\frac{\beta d^2}{\ell^2v^2}\right),
\ee
where $r$ and $b$ are the functions of $v$ defined in \eqref{PD-D20}.  Substituting (\ref{PD-D20}) into (\ref{PD-D19}) allows us to write this explicitly as
 \be\label{PD-D21}
\mathsf{F} = f_0+\frac{\mu d}{4} \left(1-\displaystyle\frac{1-v^2- \beta  }
{\sqrt{\left( 1-v^2 + \beta   \right)^2+ 4 \beta   \, v^2}}\right).
\ee
Given the value of the force $\mathsf{F}$, the {\it kinetic relation} \eqref{PD-D21} gives the value of the propagation speed\footnote{Alternatively the kinetic relation follows by enforcing the requirement \eqref{20170928-eq10} on the solution \eqref{20170926-eq1}. } $v$. {{Observe that  $\mathsf{F} > f_0$ and therefore that $(\mathsf{F} - f_0) v \geq 0$.}}

\begin{figure}[h]
\centerline{\includegraphics[scale=0.4]{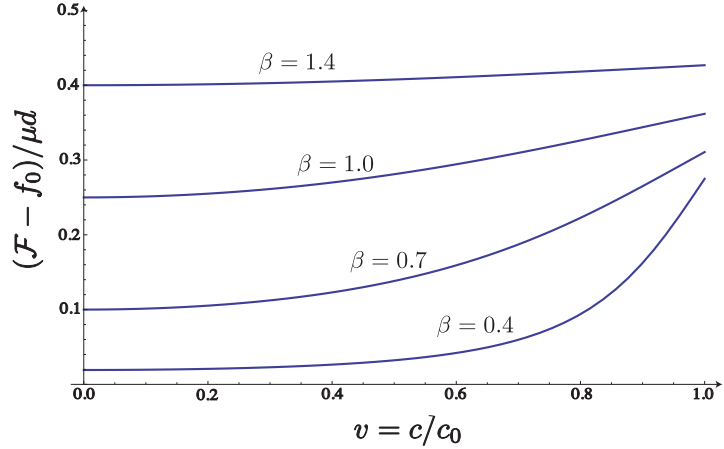}}
\caption{\footnotesize  Kinetic relation: applied force $({\mathsf{F}} - f_0)/\mu$ versus propagation speed $v = c/c_0$ for subsonic speeds at different values of $\beta = \ell^2 \mu/E$. See Figure \ref{Fig-SingleKinReln2} for a plot showing both subsonic and supersonic speeds.}
\label{20170826-fig-kinetics}
\end{figure}

Figure \ref{20170826-fig-kinetics} shows graphs of the kinetic law \eqref{PD-D21} for subsonic speeds at different values of  $\beta =\mu\ell^2/E$.  The kinetic law is shown for both subsonic and supersonic speeds  (at a single small value of $\beta  $) in Figure \ref{Fig-SingleKinReln2}.

When the lattice length scale $d$ is much smaller than the peridynamic length scale $\ell$,  we let $\beta \rightarrow \infty$ in \eqref{PD-D21} to see that $\mathsf{F} \to   f_M (= f_0 + \frac 12 \mu d )$ at each fixed $v$. On the other hand when the peridynamic length scale $\ell$ is much smaller than the lattice length scale $d$, letting $\beta \rightarrow 0$ shows that $\mathsf{F} \to f_0$ for $v<1$ and ${\mathsf{F}} \to f_M$ for $v>1$; see Figure \ref{Fig-SingleKinReln2}. Note that ${\mathsf{F}} \to f_M$ when $v\to \infty$ at fixed $\beta$.

%%%
%\begin{figure}[h]
%\centerline{\includegraphics[scale=0.3]{Fig-SingleKinRelnNew.eps}}
%\caption{\footnotesize  Kinetic relation: according to \eqref{PD-D21}, as $\beta \to 0$ the applied force ${\mathsf{F}} \to f_0$ for $v <1$ and ${\mathsf{F}} \to f_M$ for $v >1$.}
%\label{Fig-SingleKinReln}
%\end{figure}
%%%

Observe from Figure \ref{20170826-fig-kinetics} that there is a minimum value of force  necessary for kink propagation. Its value $\mathsf{F}_{\rm min}$ is found by setting $v=0$ in \eqref{PD-D21}:
 \be\label{PD-D22}
\mathsf{F}_{\rm min}= f_0+\frac{\mu d}{2}\frac{1}{1+1/\beta  }
.\ee

Finally, recall  that a kink can be in equilibrium if the applied force lies in the range given in \eqref{20170823-e9}. The rightmost expression there, i.e. the maximum value of $\mathsf{F}$ that allows for an equilibrium kink, is precisely
$\mathsf{F}_{\rm min}$.
Thus if the value of the force $\mathsf{F}$ is monotonically increased from zero, the kink will remain stationary for $\mathsf{F} < \mathsf{F}_{\rm min}$ and start to propagate once the force exceeds $\mathsf{F}_{\rm min}$.
Since $\mathsf{F}_{\rm min} > f_0$ it follows that the kink will remain in equilibrium even if the applied force exceeds the Maxwell force $f_0$ provided it is less than $\mathsf{F}_{\rm min}$ -- the phenomenon sometimes referred to as  ``lattice trapping''.

%%%%%%%%%%%%%%%%%%%%%%%%%%%%%%%%%%

\subsection{Dynamic solution when $v\to 0$.} \label{20170915-sec1}

Recall that the displacement field in the dynamic solution \eqref{20170823-PD-D18}, \eqref{PD-D21} is continuous at $z=0$, but is discontinuous at $x=0$ in the static solution \eqref{20160823-eq2}.  In order to understand this difference, we now examine the dynamic solution at small values of the speed\footnote{One can see from \eqref{20170831-eq10}$_2$ and \eqref{20170831-eq11}$_1$ that, at any fixed negative value of $z$, the limit of $u(z)$ as $v \to 0$ does not exist in the usual sense. It must be interpreted in the sense of Young measures.} $v$. It is straightforward to show from  \eqref{PD-D20} and \eqref{20170823-PD-D18} that for small $v$ the dynamic solution takes the form
\be\label{20170831-eq10}
u(z)  \sim \left\{
\ba{lll}
       \displaystyle
       u^++ \frac d2 \, \frac{1}{1 + \beta} \, e^{-b z/d},  &  z>0, \\[2ex]
        \displaystyle
       u^- - \frac d2 \, \frac{1}{1 + \beta} \,e^{b z/d} - d \frac{\beta}{1+\beta} \, \cos \left(\frac{rz}{d}\right),  &  z<0,
   \ea\right. \qquad ({\rm for \ small} \ v),
   \ee
to leading order, where from \eqref{PD-D20}
\be\label{20170831-eq11}
r \sim \frac{d}{\ell} \frac{\sqrt{1 + \beta}}{v}, \qquad b \sim \frac{d}{\ell} \frac{1}{\sqrt{1 + 1/\beta}}, \qquad ({\rm for \ small} \ v).
\ee
The small-$v$ dynamic solution \eqref{20170831-eq10}, \eqref{20170831-eq11} would be identical to the static solution \eqref{20160823-eq2}, \eqref{20160827-eq11} {\it if not for} the cosine term in \eqref{20170831-eq10}$_2$.  At fixed $v >0$ (no matter how small) $u(z)$ given by \eqref{20170831-eq10} is continuous at $z=0$.  However as $v \to 0$ observe that the frequency of oscillation of the cosine term goes to infinity whereas its amplitude remains strictly positive and finite.  Observe also that the coefficient of the cosine term
term is precisely equal to the jump in displacement $\ljump u \rjump$ of the static solution as given in \eqref{20170718-eq4}.  Of course the static solution \eqref{20160823-eq2} holds for all values of the force $\mathsf{F}$ in the range \eqref{20170823-e9} whereas the dynamic solution for small $v$ holds only for $\mathsf{F} \sim \mathsf{F}_{\rm min}$.

%%%%%%%%%%%%%%%%%%%%%%%%%%%%%%%%%%%%%%%%%%%%%%
%%%%%%%%%%%%%%%%%%%%%%%%%%%%%%%%%%%%%%%%%%%%%%

\section{Equivalency of peridynamic and certain other continuum models.}\label{20170915-sec-last}

In the preceding section we analyzed the kink propagation problem within the peridynamic theory, a problem that has been studied previously within the quasi-continuum theory by Kresse and Truskinovsky \cite{Kresse2003} and Truskinovsky and Vainchtein \cite{Trusk2006}.  It turns out that the displacement field $u(z)$ and kinetic relation $\mathsf{F}= {\mathsf{F}}(v)$ according to that theory have the same forms as the respective equations \eqref{20170823-PD-D18} and \eqref{PD-D21}  of the peridynamic theory. In fact, one can show that they coincide {\it exactly} when the peridynamic length scale has the particular value $\ell = d/\sqrt{12}$, indicating that the quasi-continuum theory can be viewed as a special case of the peridynamic theory for a particular value of the length scale, at least in the context of kink propagation.  In general, the presence of the parameter $\ell/d$ in the peridynamic theory gives it an additional degree of freedom, and its value can be chosen, for example, to match features of the discrete dispersion relation as in Figure \ref{20170708-fig-dispersionreln-new}.  Even so, the aforementioned coincidence of results when $\ell = d/\sqrt{12}$ is rather striking.

We did not anticipate this result, and it prompted us to ask the following more general question: given a  continuum model described by some differential equation, does there exist an ``equivalent'' peridynamic model?  This question is stated more precisely and investigated in the present section. The discussion pertains to any dynamical motion on an infinite interval (not only steady kink propagation).  Moreover,  {\it we will not pick a particular peridynamic micromodulus function $C$ a priori}.  Instead, we shall {\it determine} the function $C$ that makes the peridynamic model equivalent to the other model described by a  differential equation.  This will be illustrated  using two explicit examples  (the Boussinesq and quasi-continuum models) after a more general analysis.

We shall use the notation $\mathscr{F}\{g\}$ for the Fourier transform of a function $g(x,t)$. Since different authors use slightly different definitions of this, we note that the one we use is
\be\label{20171031-eq1}
G(k,t) = \mathscr{F}\{g\} =  \frac{1}{2\pi} \int_{-\infty}^\infty g(x,t) {\rm e}^{ikx}\, dx , \qquad
g(x,t) = \mathscr{F}^{-1}\{G\} = \int_{-\infty}^\infty G(k,t) {\rm e}^{-ikx}\, dk .
\ee

Our strategy is based on Whitham \cite{Whitham1999}: we first Fourier transform the peridynamic equation of motion. The result involves the acoustic function $M(k)$ which can be determined from \eqref{20170715-eq10} if we know the micromodulus function $C(x)$.  We invert this relation and obtain a formula for calculating $C(x)$ when $M(k)$ is known. Then, given some other  continuum model described by a  differential equation, if the Fourier transform of that  equation of motion has the same form as the Fourier transformed peridynamic equation of motion for some $M(k)$ we say the models are equivalent, and we can calculate the associated micromodulus function  $C(x)$ provided that a certain integral exists.

We start with the peridynamic equation of motion
\be\label{20170828-eq16}
\int_{-\infty}^\infty C(\xi) \big[ u(x+\xi,t) - u(x,t)\big]\, d\xi \  + \  b(u) = \rho   {u}_{tt},
\ee
where the body force $b(u)$ may equal $\mathsf{F} - w'(u)$ as in the preceding section. While one could consider more general body forces, e.g. $b(u,u_x)$, one of the attractive features of peridynamic theory is that it involves no spatial derivatives of $u$ and so including a $u_x$-dependency in the body force would be counter to that spirit. Therefore we limit attention to body forces of the form $b=b(u)$. Fourier transforming \eqref{20170828-eq16}
leads to
\be\label{20170828-eq21}
- M(k) \mathscr{F}\{u\}  + \  \mathscr{F}\big\{ b(u) - \rho   {u}_{tt} \big\} = 0,
\ee
where the acoustic function $M(k)$ is
\be\label{20170828-eq15}
M(k) = \int_{-\infty}^{\infty} C(\xi) \big[1-  \e^{ik\xi} \big] \, d\xi .
\ee
Given the micromodulus function $C(x)$, \eqref{20170828-eq15} is an equation for calculating the acoustic function $M(k)$.  Our immediate goal is to invert this relation to calculate $C(x)$ in terms of $M(k)$.

Observe that for continuous displacement fields $u(x)$,
$$
\int_{-\infty}^\infty \delta(\xi)[u(x+\xi) - u(x)] d \xi = 0.
$$
Therefore, adding a Dirac $\delta$-function to the micromodulus function $C$ does not change the peridynamic equation of motion \eqref{20170828-eq16}.  Likewise it does not change the acoustic function $M(k)$ given by \eqref{20170828-eq15}. Thus we will only determine $C$ to within a $\delta$-function.

First, Dayal \cite{Dayal2017} has shown that if $M(k)$ has a finite limit $M_\infty$ as $k \to \infty$ then
\be\label{20171031-eq3}
M_\infty  \coloneqq \int_{-\infty}^{\infty} C(\xi) \, d\xi.
\ee
Thus we can write \eqref{20170828-eq15} as
\be
M(k) - M_\infty =  - \int_{-\infty}^{\infty} C(\xi)  \e^{ik\xi}  \, d\xi ,
\ee
which says that $-C(k)$ is the Fourier transform of $M(k) - M_\infty$. Thus by using the inverse Fourier transform \eqref{20171031-eq1} we obtain
\be \label{20170828-eq22}
C(x) = - \frac{1}{2\pi} \int_{-\infty}^{\infty} (M(k)-M_\infty)  \e^{-ikx}  \, dk,
\ee
which gives the micromodulus function $C$ corresponding to the acoustic function $M$.

The preceding result is not valid if $M(k) \to \infty$ as $k\to \infty$.  Suppose that $M(k) = O(k^{2n})$ as $k \to \infty$ for some integer $n \geq 1$, with
\be \label{20171102-eq1}
\frac{M(k)}{k^{2n}} \to \mathsf{m}_\infty \qquad {\rm as} \  k \to \infty,
\ee
$\mathsf{m}_\infty$ being a finite number.  In general, the micromodulus function will then involve a generalized function and its derivatives as in Whitham \cite{Whitham1999}. Even so a formal characterization of $C$ is possible as follows: define the function $\mathscr{C}(x)$ by
\be\label{20171102-eq2}
\mathscr{C}(x) \coloneqq  \ - \frac{(-1)^n}{2 \pi } \int_{-\infty}^\infty  \left( \frac{M(k)}{k^{2n}} - \mathsf{m}_\infty\right) {\rm e}^{-ikx}\, dk - (-1)^n \mathsf{m}_\infty\delta(x).
\ee
One can show by taking the Fourier transform of this equation and integrating by parts that
\be\label{20171102-eq3}
C(x) = \mathscr{C}^{(2n)}(x)
\ee
obeys \eqref{20170828-eq15}, and therefore that \eqref{20171102-eq3} can be taken to be the micromodulus function in this case. Here $\mathscr{C}^{(k)}$ denotes the $k$th derivative of $\mathscr{C}$.
When $n=0$, \eqref{20171102-eq2}, \eqref{20171102-eq3} coincides with \eqref{20170828-eq22} to within a $\delta$-function.

Given a continuum model  based  on some differential equation,  if the Fourier transform of the associated equation of motion has the form
\eqref{20170828-eq21} for some $M(k)$,  we say the models are equivalent, and one can determine the micromodulus function $C$ of the equivalent peridynamic model through either \eqref{20170828-eq22} or \eqref{20171102-eq2}, \eqref{20171102-eq3}, provided the relevant integrals exist.

%%%%%%%%%%%%%%%%%%%%%%%%%%%%%%%%%%%%%%%%

In the next subsection two explicit differential-equation-based continuum models stemming from the basic lattice equation of motion \eqref{20170826-gerol-eq5} will be examined.
They will both be of the generic form
\be\label{(i)}
{\cal L}_1 u_{xx} + {\cal L}_2 [b(u) - \rho u_{tt}] = 0,
\ee
where the differential operators ${\cal L}_1$ and ${\cal L}_2$ are
\be\label{(i)derma}
{\cal L}_1 = \sum_{n=1}^{m_1}  a_{2n-2} \frac{\partial^{2n-2}}{\partial x^{2n-2}}, \qquad
{\cal L}_2 = \sum_{n=1}^{m_2} b_{2n-2} \, \frac{\partial^{2n-2}}{\partial x^{2n-2}} ,
\ee
for some constant $a_i$'s and $b_i$'s, and integers $m_1 \geq 1, m_2 \geq 1$. Recalling that for any function $g$
$$
\mathscr{F}\left\{\frac{\partial^ng}{\partial x^n}\right\} = (-ik)^n\mathscr{F}\{g\},
$$
we take the Fourier transform of \eqref{(i)}, \eqref{(i)derma} to get
$$
q(k) \calF\{u\} + p(k) \calF\{b(u) - \rho u_{tt}\} = 0,
$$
where the polynomials $q(k)$ and $p(k)$ are
\be\label{(iii)}
\displaystyle q(k) =  \sum_{n=1}^{m_1} a_{2n-2}  (-1)^{n} k^{2n}, \qquad
\displaystyle p(k) =  \sum_{n=1}^{m_2} b_{2n-2}  (-1)^{n-1} k^{2n-2} .
\ee
Observe now that the Fourier transform of the differential equation \eqref{(i)} is identical to that of the peridynamic equation of motion \eqref{20170828-eq16} provided one takes
\be \label{20171031-eq2}
M(k) = - \frac{q(k)}{p(k)}.
\ee
The micromodulus function of the peridynamic model equivalent to the differential equation model  \eqref{(i)} is then given by either \eqref{20170828-eq22} or \eqref{20171102-eq2}, \eqref{20171102-eq3}.

%%%%%%%%%%%%%%%%%%%%%%%%%%%%%

\subsection{Two examples.}\label{20180224-sec5}

We now illustrate the preceding result with two explicit examples. Returning to the discrete model  \eqref{20170826-gerol-eq5} and considering nearest neighbor interactions only, we obtain the following delay-differential equation to be satisfied by the displacement field $u(x,t)$:
\be\label{gerol-eq5}
E \, \frac{u(x+d,t) - 2 u(x,t) + u(x-d, t)}{d^2}  +  b(u) = \rho   {u}_{tt}(x,t)  ,
\ee
where we have set  $E = \kappa_1$ and $b(u) = \mathsf{F} - w'(u)$. Suppose that at each $t$, the displacement field $u(x \pm d,t)$ admits a Taylor series representation
\be \label{gerol-eq6}
u(x \pm d,t)
=  \sum_{n=0}^\infty  \frac{(\pm d)^{n}}{n!} \frac{\partial^n}{\partial x^n} \, u(x,t).
\ee
By using \eqref{gerol-eq6} we can write  \eqref{gerol-eq5} in the following alternative form\footnote{For the discrete model with N nearest neighbor interactions, Seleson et. al. \cite{Seleson2009} have demonstrated that a form similar to \eqref{gerol-eq7} agrees well with the higher-order gradient model obtained from peridynamics by a Taylor series expansion.}:
\be\label{gerol-eq7}
E {\cal L} u_{xx}  + b(u)   =  \rho   u_{tt}
\ee
where $\calL$ denotes the operator
\be\label{gerol-eq8}
{\cal L} = \sum_{n=1}^\infty 2  \frac{d^{2n-2}}{(2n)!} \frac{\partial^{2n-2}}{\partial x^{2n-2}} = 1 + \frac{d^2}{12} \frac{\partial^{2}}{\partial x^{2}} + \ldots .
\ee

If we keep only the first term in the series representation of ${\cal L}$, \eqref{gerol-eq7} reduces to
$$
E u_{xx}  + b(u)   =  \rho   u_{tt} .
$$
This is the equation of motion associated with {\it classical elasticity},
encountered previously in Section \ref{20170912-sec1}. If we keep two terms, we get  an equation corresponding to a particular strain-gradient theory of elasticity:
\be\label{20170916-eq2}
 \frac{Ed^2}{12} u_{xxxx}  + E u_{xx} + b(u)   =  \rho   u_{tt} .
 \ee
 This is the so-called {\it Boussinesq model} mentioned in the introduction.

Since \eqref{20170916-eq2} involves a term $u_{xxxx}$, this implies that additional boundary conditions beyond the usual ones of elasticity are needed  in order to solve an initial-boundary value problem.
In order to avoid this, Rosenau \cite{Rosenau1987} proposed first writing \eqref{gerol-eq7} as
\be\label{gerol-eq10}
E  u_{xx}  + {\cal L}^{-1} \left[b(u) -  \rho   u_{tt} \right]=0,
\ee
where
\be\label{882zz}
{\cal L}^{-1} = \sum_{n=1}^m c_{2n-2} \, \frac{\partial^{2n-2}}{\partial x^{2n-2}}  = 1 -  \frac{d^2}{12}  \frac{\partial^{2}}{\partial x^{2}}
+\frac{d^4}{240} \frac{\partial^{4}}{\partial x^{4}} + \ldots.
\ee
On keeping the first two terms in the series representation of $\calL^{-1}$,
\eqref{gerol-eq10} reduces to
\be\label{gerol-eq12}
\frac {\rho  d^2}{12}  u_{xxtt}  +   E u_{xx} - \rho   u_{tt}  + b(u) -  \frac {d^2}{12} \frac{\partial^2}{\partial x^2}  \left[ b(u) \right]
 = 0.
\ee
This is the {\it quasi-continuum approximation} referred to in the introduction.
Observe that \eqref{gerol-eq12} does not involve $x$-derivatives of $u$ of order higher than two.  The term $u_{xxtt}$ can be associated with a term $\frac 12 u_{xt}^2$ in the kinetic energy and is referred to as `micro-inertia'' in strain gradient models of continua, Mindlin \cite{Mindlin1964}.

% %%%%%%%%%%%%%%%%%%%%%%%%%%%%%%%%%%%%%%%%
%
% \subsubsection{Example: relation between the peridynamic model and a class of quasi-continuum models.} \label{20170828-subsec13}

\noindent{\bf Example 1:} Quasi-continuum model:  This is described by equation \eqref{gerol-eq12} and is the special case $m_1 = 1, m_2=2, a_0 = E, b_0 = 1$ and $b_2=-d^2/12$ of the general representation
\eqref{(i)}, \eqref{(i)derma}.
 The polynomials $q(k)$ and $p(k)$, and the dispersion function $M(k)$, are now given by \eqref{(iii)}, \eqref{20171031-eq2} and \eqref{20171031-eq3} to be
$$
q(k) = - E k^2, \qquad p(k) =  1 + \frac{d^2}{12} k^2, \qquad M(k) = - \frac{q(k)}{p(k)} =\frac{E k^2}{1 + {d^2} k^2/12},\qquad M_\infty = \frac{12 E}{d^2},
$$
and so by \eqref{20170828-eq22},  the micromodulus function of the equivalent peridynamic material is
$$
C(x) = - \frac{1}{2\pi} \int_{-\infty}^{\infty} \left( \frac{E k^2}{1 + k^2d^2/12} - \frac{12 E}{d^2} \right) e^{-{\rm{i}}kx} \, \de k  .
$$
Evaluating this integral yields (to within a $\delta$-function)
 \be\label{RL7}
C(x)=\frac{12\sqrt{3} E}{d^3}e^{-\sqrt{12}|x|/d}.
\ee
This is the micromodulus function of the peridynamic model equivalent to the quasi-continuum model. Remarkably, this micromodulus function is precisely of the form (\ref{gerol-eq14}) that we adopted in Section \ref{20170826-sec-888} (motivated there by an analysis of nonlocal effects  by Silling \cite{Silling2014}). Comparing (\ref{RL7}) with (\ref{gerol-eq14}) shows (not surprisingly anymore!) that the former is the special case of the latter corresponding to $\ell = d/\sqrt{12}$.

\

\noindent{\bf Example 2:}  Strain gradient model (Boussinesq):  This is described by \eqref{20170916-eq2}
and corresponds to the special case $m_1 = 2, m_2=1, a_0 = E, a_2 = Ed^2/12$ and $b_0 = 1$  of
\eqref{(i)}, \eqref{(i)derma}.  From \eqref{(iii)} the polynomials $p(k)$ and $q(k)$ now are
$$
q(k) = - Ek^2 + \frac{Ed^2}{12} k^4, \qquad p(k)=1.
$$
%and therefore by \eqref{(v)}  the associated micromodulus function is
%$$
%C(x) = \frac{1}{2\pi} \int_{-\infty}^\infty \Big( - Ek^2 + \frac{Ed^2}{12} k^4\Big)  e^{-ikx} dk  = E \left(\delta''(x) + \frac{d^2}{12} \delta''''(x)\right),
%$$
%where $\delta(x)$ is the Dirac $\delta$-function and primes denote derivatives.
In this case $M(k) = -q(k)/p(k) = O(k^4)$ as $k \to \infty$.  Therefore we must use \eqref{20171102-eq2} and \eqref{20171102-eq3} with $n=2$ and $\mathsf{m}_\infty =  - {Ed^2}/{12}$ to find the equivalent micromodulus function. From \eqref{20171102-eq2},
$$
\mathscr{C}(x) =  \ - \frac{1}{2 \pi } \int_{-\infty}^\infty  \frac{E}{k^2}  {\rm e}^{-ikx}\, dk  +   \frac{Ed^2}{12} \delta(x)
=  E|x|+\frac{Ed^2}{12} \delta(x),
$$
and so from \eqref{20171102-eq3},
$$
C(x) = \mathscr{C}''''(x) = E\delta''(x) +\frac{Ed^2}{12} \delta''''(x),
$$
where the primes denote derivatives. This is in fact a special case of a result in Section 11 of Whitham \cite{Whitham1999}.

In the two examples above, we effectively equated \eqref{(i)} to \eqref{gerol-eq7} by factoring the operator $\calL$ into $E \calL = {{\cal L}_2}^{-1}\calL_1$. Thus $\calL_1$ and $\calL_2$ are the Pad\'{e} approximants of $E \calL$; see also Dayal \cite{Dayal2017}. Further approximations of \eqref{gerol-eq7} can be generated in this manner.

%%%%%%%%%%%%%%%%%%%%%%%%%%%%%%%%%%%%%%%%%%%%%%
%%%%%%%%%%%%%%%%%%%%%%%%%%%%%%%%%%%%%%%%%%%%%%

\section{Conclusions.}

\parskip 0.0in

In this study we used peridynamic theory to  study the equilibrium and steady propagation of a lattice defect - a kink -- in one-dimension. Since peridynamic theory does not involve spatial gradients of the displacement field, it is particularly well suited for studying defect propagation.

A material in the peridynamic model is characterized by a micromodulus function. The specific micromodulus function we used involved two parameters, a modulus and a length scale.  The values of these two parameters can be chosen by matching features of the dispersion relations of the peridynamic and discrete lattice models at both short and long  wavelengths.

As a kink propagates, it progressively transforms the material from one state to another and this was captured in the model by a double-well potential. The material is in one (pre-transformed) state ahead of the kink and in another (transformed) state behind it.  The material transforms locally as the propagating kink passes through each point.

Though the peridynamic operator was linear in our formulation, the equation of motion was nonlinear due to the presence of the double-well potential in the body force term.  Even so, we were able to study the problem analytically, primarily by taking advantage of a convenient form of the double-well potential.

We found that the kink cannot propagate if the applied force is less than a certain critical value $\mathsf{F}_{\rm min}$.  For values of force exceeding $\mathsf{F}_{\rm min}$, the kink can propagate at a steady speed and there is a relation $\mathsf{F} = \mathsf{F}(v)$ between the applied force  and propagation speed. We determined
$\mathsf{F}_{\rm min}$ and $\mathsf{F}(v)$.

\parskip 0.1in

A second contribution of this study is that we showed that the dynamical solutions of certain differential-equation-based models of a continuum are the same as those of the peridynamic model provided the micromodulus functions are chosen suitably.  We derived a formula for calculating the micromodulus function of the equivalent peridynamic model.  This result holds for all dynamical processes and is not limited to steady kink propagation.
This ability to replace a differential-equation-based model with an equivalent peridynamic one may prove useful when numerically studying more complicated problems such as those involving multiple and interacting defects.

{ However there are several open questions.  As one reviewer asked, is there a way to directly convert the quasi-continuum partial differential equation into the peridynamic integral equation without the use of transforms? Moreover, the equivalence between the quasi-continuum  and peridynamic {equations} demonstrated in this paper does not address the role of boundary conditions. Are these models completely equivalent particularly on bounded domains? We do not know the answers to these important questions at this time.

Finally, as noted in Section \ref{20170915-sec-last}, the two examples considered there correspond to two distinct classes of acoustic functions $M(k)$. The acoustic function in Example 1 has the property that $M(k)$ is bounded as $k \to \infty$. On the other hand in Example 2, $M(k)$ is not bounded in this limit though $M(k)/k^{2n}$ is bounded for some $n\geq 1$.  In the latter case, the acoustic function involves generalized functions and the peridynamic kernel loses its nonlocality, a core feature of the peridynamic theory. It is natural then to inquire as to the class of all quasi-continuum models that yield truly nonlocal kernels. It maybe seen from
\eqref{(i)} - \eqref{20171031-eq2} and \eqref{882zz} that the acoustic function associated with the general quasi-continuum equation  \eqref{gerol-eq10} can be written as
$$
M(k) = - \frac{q(k)}{p(k)} \qquad {\rm where} \ q(k) = - Ek^2, \quad  p(k)=\sum\limits_{n=1}^{m_2}c_{2n-2}(-1)^{(n-1)}k^{(2n-2)}.
$$
Therefore $M(k)$ is bounded as $k \to \infty$ for any integer $m_2 \geq 2$ and so the quasi-continuum equation
 \eqref{gerol-eq10} leads to truly nonlocal, square integrable peridynamic kernels $C$ for any number of terms $\geq 2$ in the representation \eqref{882zz}. This is not true of the generalized Boussinesq model
 \eqref{gerol-eq7},  \eqref{gerol-eq8}.
}

%%%%%%%%%%%%%%%%%%%%%%%%%%%%%%%%%%%%%%%%%%%%%%%
%%%%%%%%%%%%%%%%%%%%%%%%%%%%%%%%%%%%%%%%%%%%%%%

\noindent{\Large \bf Acknowledgement.}

\noindent The authors express their appreciation to Professor Kaushik Dayal for his interest in this work and  for his insightful comments on an earlier draft of this manuscript.  {{The authors are also grateful to the reviewers whose comments have led to valuable additions to this paper.}}  L. Wang acknowledges the support of the National Natural Science Foundation of China under Grant 11521202, and the Chinese Scholarship Council.

%%%%%%%%%%%%%%%%%%%%%%%%%%%%%%%%%%%%%%%%%%%%%
%%%%%%%%%%%%%%%%%%%%%%%%%%%%%%%%%%%%%%%%%%%%%%

%\begin{spacing}{1.0}

%\end{spacing}

\clearpage

%%%%%
\begin{figure}[h]
\centerline{\includegraphics[scale=0.5]{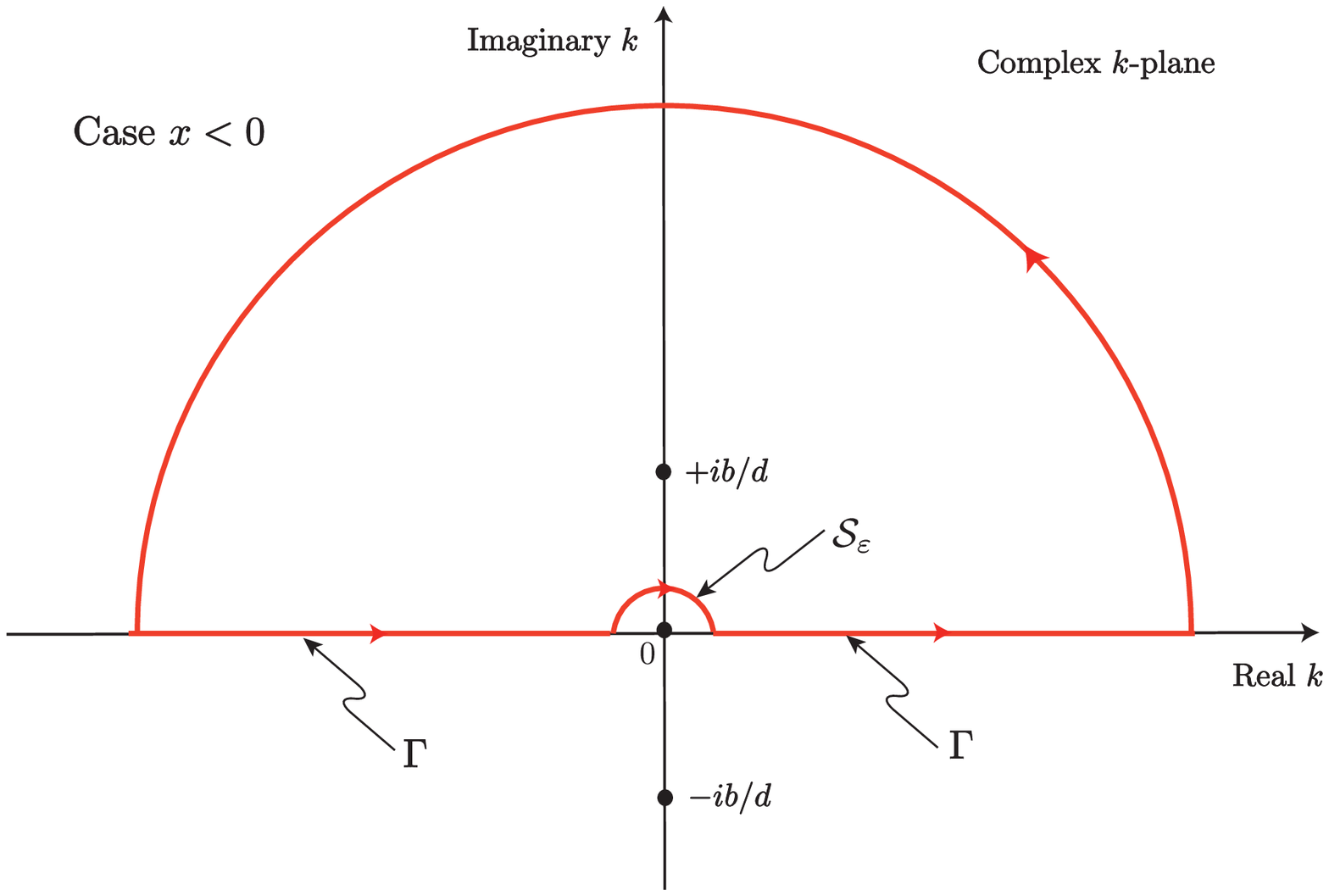}}
\caption{\footnotesize  Stationary kink: The complex $k$-plane showing poles and integration path. The figure has been drawn for the case $x<0$. When $x > 0$ the large semi-circle must be taken in the lower half-plane.}
\label{fig-Fig-ComplexPlane1}
\end{figure}
%%%%%

%%%%%
\begin{figure}[h]
\centerline{\includegraphics[scale=0.5]{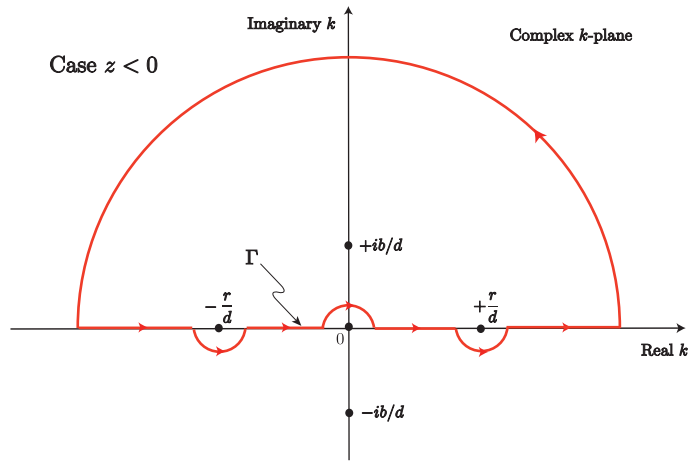}}
\caption{\footnotesize  Propagating kink: The complex $k$-plane showing poles and integration path. }
\label{fig-Fig-ComplexPlane2}
\end{figure}
%%%%%

\end{document}